\documentclass[a4paper,fleqn,usenatbib,useAMS]{mnras}

\usepackage{graphicx}	
\usepackage{amsmath}	
\usepackage{amssymb}	
\usepackage{multicol}        
\usepackage{bm}		
\usepackage{pdflscape}	
\usepackage{tabularx}
\usepackage{adjustbox}

\usepackage[T1]{fontenc}
\usepackage{ae,aecompl}

\usepackage{txfonts}

\title[Type-C QPOs in soft state]{\textit{Constraining black hole spins with low-frequency quasi-periodic oscillations in soft states}}

\author[Franchini, Motta and Lodato]{ Alessia Franchini$^{1}$\thanks{alessia.franchini@unimi.it}, Sara Elisa Motta$^2$ and Giuseppe Lodato $^1$\\
$^1$Dipartimento di Fisica, Universit\`a Degli Studi di Milano, Via Celoria, 16, Milano, I-20133, Italy\\
$^2$University of Oxford, Department of Physics, Astrophysics, Denis Wilkinson Building, Keble Road, OX1 3RH, Oxford , UK
}

\date{Last updated ...; in original form ...}

\pubyear{2016}

\begin{document}
\label{firstpage}
\pagerange{\pageref{firstpage}--\pageref{lastpage}}
\maketitle

\begin{abstract}

Black hole X-ray transients show a variety of state transitions during their outburst phases, characterized by changes in their spectral and timing properties. In particular, power density spectra (PDS) show quasi periodic oscillations (QPOs) that can be related to the accretion regime of the source. 
We looked for type-C QPOs in the disc-dominated state (i.e. the high soft state) and in the ultra-luminous state in the RXTE archival data of 12 transient black hole X-ray binaries known to show QPOs during their outbursts. 
We detected 6 significant QPOs in the soft state that can be classified as type-C QPOs. 
Under the assumption that the accretion disc in disc-dominated states extends down or close to the innermost stable circular orbit (ISCO) and that type-C QPOs would arise at the inner edge of the accretion flow, we use the relativistic precession model (RPM) to place constraints on the black hole spin.
We were able to place lower limits on the spin value for all the 12 sources of our sample while we could place also an upper limit on the spin for 5 sources.
\end{abstract}




\begingroup
\let\clearpage\relax
\endgroup
\newpage

\section{Introduction}

Black hole (BH) low-mass X-ray binaries (LMXBs) are double systems harboring a stellar mass black hole that accretes matter from a Sun-like companion star.
The vast majority of BH LMXBs are transient systems, i.e. they spend most of their life in a dim, quiescent state (L$_X$ $\sim$ 10$^{30}$--10$^{34}$ erg/s), occasionally interrupted by bright outbursts during which they show large increases in luminosity (reaching occasionally the Eddington limit) and remarkable variations in their spectral and fast variability properties \citep{Belloni2016}.
These changes can be described through the Hardness-Intensity Diagram (HID) \citep{Homan2001,Belloni2011}, where hysteresis loops can be identified \citep{Miyamoto1992} and different branches of a q-shaped track roughly correspond to different spectral/timing states. Most active BH LMXBs show five main different states: the Low Hard State (LHS), dominated by Comptonized emission;  the High Soft State (HSS), where the thermal emission from an accretion disc dominates the spectrum; the Hard Intermediate State (HIMS) and the Soft Intermediate State (SIMS), where the spectrum shows the contributions from both the accretion disc and the Comptonized emission, but where the most dramatic changes in the timing properties are observed. A few source have also shown an {\it anomalous} or ultra-luminous state (ULS), which can be seen as a soft-intermediate and hard-intermediate state in which the luminosity is significantly higher with respect to the other states (see, e.g., \citealt{Motta2012}).
While the hard states are characterized by higher variability ($30-40$\% rms in the LHS and $5-20$\% rms in the intermediate states), the variability in the ULS and in the HSS is very low (around and below $5$\% rms, respectively).

The different states are defined based on the spectral properties of the source and on the inspection of the power density spectra (PDS), where changes in the fast time variability can be clearly observed. 
The most remarkable features detected in the PDS are narrow peaks known as quasi-periodic oscillations (QPOs), whose centroid frequency are thought to be associated with dynamical time scales in the accretion flow. 
The origin of QPOs is still under debate and several models have been proposed to explain their existence (see \citealt{Motta2016} for a review). However, it is commonly accepted that most QPOs originate in the vicinity of the black hole, thus they potentially carry information on the properties and motion of matter in a strong gravitational field.

QPOs are usually divided in two groups based on their frequency: low frequency  and high frequency (LFQPOs and HFQPOs, respectively). LFQPOs have frequencies typically varying from a few mHz up to $\sim 30$ Hz and are further divided into three types - type-A, type-B and type-C - based on their properties (e.g. peak frequency, width, energy dependence and phase lags) and on the associated broad band noise in the PDS (shape and total variability level, \citealt{Wijnands1999}, \citealt{Casella2005}, \citealt{Motta2011}).  According to the classification of \cite{Homan2005a},  Type-A QPOs are commonly found in the SIMS  and are characterized by a weak and broad peak around $6-8$ Hz. \footnote{However, we note that more recent works \citep{Belloni2011,Motta2016,Belloni2016} placed this type of QPOs in the HSS, since they typically appear at total fractional rms below 5\% in BHBs.} Their origin is still under debate. 
Type-B are usually seen in the SIMS (which is defined by their presence, \citealt{Belloni2016}). They are characterized by a relatively strong and narrow peak around $6$ Hz \citep{Motta2011} and it has been suggested that they are associated with relativistic jets during the transition from the hard to the soft state \citep{Fender2009}.
Type-C QPOs are found essentially in any spectra/timing state  (including the ULS, \citealt{Motta2016}). They are characterized by a strong (up to $20$\% rms), narrow ($\nu/\Delta \nu \geq 3$, but often much larger) and variable peak (its frequency and strength varying by few orders of magnitude along an outburst; see, e.g., \citeauthor{Motta2015} \citeyear{Motta2015}) superposed on a flat-top noise that steepens above a frequency comparable to the centroid frequency of the QPO.
Type-C QPOs appear at low frequencies (from a few mHz up to tenths of Hz \citealt{Motta2011}) in the hard state where the disc is thought to be truncated at a large radius (tens to hundreds of R$_{g}$). They increase their frequency  during the transition from the hard to the soft state, reaching their maximum in the HSS, where they show frequencies around 30 Hz (\citealt{Motta2014}, \citealt{Motta2014a}).
HFQPOs are fairly rare in BH binaries (while they are commonly observed in NS systems, where they are referred to as kHz QPOs, see \citealt{VDK2006}) and have been detected only in a few sources \citep{Belloni2012}. HFQPOs seem to appear preferentially in the HSS and in the ULS, even though observational biases could be at play because of the low number of detections currently available.  
\cite{Remillard2006}, using a different classification (Table 1 in \citealt{McClintock2009}), showed that most HFQPOs have been detected in observations with a substantial power-law contribution.

\smallskip

There are different models that attempt to describe the origin of QPOs in LMXBs. The relativistic precession model (RPM) is one of those. This model has been originally proposed by \cite{Stella1998,Stella1999} and was recently revisited by \cite{Motta2014,Motta2014a,Ingram2014}, to explain three types of QPOs: type-C QPOs, the lower and the upper HFQPOs.
In the RPM, the centroid frequency of three different types of QPOs is associated with the  frequencies of a test particle orbiting a spinning black hole, as predicted by the theory of General Relativity.
The Lense-Thirring (LT) frequency is associated with the type-C QPO, the periastron precession frequency with the lower HFQPO and the orbital motion corresponds to the upper HFQPO. The RPM predicts that the expected values of these frequencies depend solely on the fundamental parameters of the BH, i.e. mass and spin, and on the radius at which the particle motion occurs (i.e. where the QPOs are produced). Based on this fact, once a type-C QPO and two HFQPOs are observed simultaneously, one can obtain a precise measurement of the fundamental parameters of the black hole, spin and mass \citep{Motta2014}. A spin measurement can also be obtained when only two simultaneous QPOs of the relevant types are detected and when an independent mass measurement, for instance determined dynamically, is available \citep{Motta2014a}. If the mass is unknown, the RPM still allows to place limits on the mass and spin of the black hole (see \citealt{Ingram2014}).

\cite{Ingram2009} proposed a model based on relativistic precession that requires a geometrically thin accretion disc \citep{Shakura1973} truncated at some radius (the inner disc truncation radius) filled in with an inner hot accretion flow that is able to rigidly precess around the compact object spin axis. According to this model type-C QPOs are produced by the Lense-Thirring precession of a radially extended portion of the hot inner accretion flow that produces a modulation of the X-ray flux, differently from the RPM where a test particle is responsible for the QPO.
In the first case, the QPO frequency is a weighted average of the free precession frequencies at the various radii. However, for the disc to precess rigidly, its radial extent must be narrow  (tens of $R_{\rm g}$) \citep{LodatoF2013,Franchini2016,Nixon2016}.
In particular, in the HSS, where the accretion disc inner radius is expected to be close to or coincident with the innermost stable circular orbit (ISCO, e.g. \citealt{Dunn2010}), the inner hot (geometrically thick) accretion flow is expected to be extremely narrow (its outer radius determined by the inner thin disc truncation radius). Therefore, its precession frequency is reasonably well described by that of a test particle orbiting at the ISCO. Thus, in the HSS and in particular close to the ISCO, the RPM and the rigid precession are practically coincident and the RPM represents a good approximation of the rigid precession \citep{Ingram2014}.

The measurement of the black hole parameters is a major issue in astrophysics. While the mass can be estimated through often complex dynamical studies, the spin can only be inferred in an indirect way. In the HSS measurements of the spins can be obtained through X-ray spectroscopy by modelling the thermal disc emission in the spectrum \citep{McClintock2011,McClintock2014,Steiner2010,Steiner2011,Steiner2012}. In the hard and intermediate states, the spin is instead measured by modelling the disc reflection spectrum, in particular by modelling the effects of the spin on the Iron-K$\alpha$ line \citep{Miller2007,Reynolds2013}. Both methods
rely upon identifying the inner radius of the accretion disc with
the ISCO. In addition, for the spectral continuum
method, one must also know the mass of the black hole,
the inclination of the accretion disc (generally assumed equal to
the inclination of the binary system) with respect to the line of sight and the distance to the binary. For these reasons, such methods are often affected by large systematic errors and in a few cases they have lead to inconsistent values of the black hole spin when used on the same source \citep{Shafee2006}. The RPM, on the other hand, is based only on measurements that can be inferred through very precise X-ray timing measurements (i.e. the precision of the measurement of a QPO centroid frequency is only limited by the time resolution of the data used, which is usually extremely high). Therefore, the RPM can be used to estimate BH spins independently from other methods. 

In this work we apply the RPM to QPOs detected in the HSS, i.e. produced close or at the ISCO, placing limits on the black hole spin of a number of sources by making reasonable assumptions on the mass, when this is not known \textit{a priori}.


\section{Observations and data analysis}\label{sec:obs_anal}

Based on \cite{Motta2015} we analyzed {\it RXTE}/Proportional Counter Array (PCA) archival observations of $12$ transient BHBs showing QPOs. We selected those sources that show a standard HSS among the $14$ sources considered by  \cite{Motta2015}. We did also include observations in the ULS for those sources that showed this kind of state during their outbursts. In this work, we excluded {\it Swift} J1753.5-0127 since it has not shown a HSS in the {\it RXTE} era \citep{Soleri2013},  and XTE J1720-318, which did not show QPOs during the outburst observed by {\it RXTE}.
We looked for QPOs in the HSS and in the ULS, where they are thought to be produced at or close to the ISCO (see \citealt{Motta2012}, \citealt{Motta2015}). We analyzed all the observations in the {\it RXTE} catalogue in the HSS and ULS that have not been included in \cite{Motta2014,Motta2014a,Motta2015}. Then we added also the QPOs found in the previous works.

We computed the PDS using a custom software under \textsc{IDL} in the total and hard energy band ($\sim 2-37$ keV,  absolute channels 0-64, and $\sim 6-37$ keV, absolute channels 14-64, respectively). Since QPOs typically show a hard spectrum \citep{Sobolewska2006}, producing the PDS using only hard photons maximizes the chances of detecting a faint signal, whose intrinsic variability could be diluted by the non-variable photons from the accretion disc, that dominates in the soft states and emits mostly below $6$ keV. 
For similar reasons, we did not consider the photons above channel $64$: the efficiency of the PCA drops dramatically above $\sim$30 keV and the variability coming above this energy is mostly Poisson noise. In addition, in the HSS the energy spectrum is typically very steep, with very little or no signal above $\sim$20 keV.
In a few cases we could not produce PDS in the hard band since the data modes did not allow it, therefore we only analyzed the total energy band PDS (channel 0-64 if available, otherwise channels 0-249).
We used $16$ s-long intervals and a Nyquist frequency of $2048$ Hz to produce the PDS, which were then normalized following \cite{Leahy1983} and converted to square fractional rms \citep{Belloni1990}. We measured the integrated fractional rms  (i.e. the rms from the entire PDS) in the 2-15 keV band and between 0.1 and 64 Hz. From now on, if not otherwise stated, by rms we refer to the integrated fractional rms.

Then, we inspected all the PDS selecting only those where we detected one or more narrow peaks and fitted them using the standard XSPEC fitting package with a one-to-one energy-frequency conversion and a unity response matrix.
Following \cite{Belloni2002}, we fitted the noise components with a number of broad Lorentzians and the hints of QPOs with narrow Lorentzians \footnote{Type-C QPOs sometimes show significant harmonic content especially in the hard states. However, a single Lorentzian is always enough to fit type-C QPOs in the HSS.}. A constant component was added to take into account the contribution of the Poisson noise.
We estimated the statistical significance of these peaks and we excluded all features with single trial significance below $3\sigma$.
Then we considered also the number of trials in the calculation of the significance and we discarded all the peaks that resulted not significant after this analysis.
We rejected also the peaks with quality factor (i.e. the ratio between centroid frequency and FWHM of the QPO peak) $Q<2$.
Since in the HSS both type-C QPOs and HFQPOs can be seen, we classified the QPOs we found following \cite{Motta2011} and \cite{Motta2012}.

\section{The relativistic precession model: constraining the spin}\label{sec:method}

The fundamental frequencies of a particle orbiting a spinning black hole are given by (from lowest to highest frequency):

\begin{equation}
\nu_{\mathrm{LT}} = \nu_{\phi} \left[ 1 - \left(1 \mp 4ar^{-3/2} +3a^2r^{-2}\right)^{1/2}\right] \label{eq:nod}
\end{equation}

\begin{equation}
\nu_{\mathrm{per}} = \nu_{\phi} \left[ 1 - \left(1-6r^{-1} \pm 8ar^{-3/2} -3a^2r^{-2}\right)^{1/2}\right] \label{eq:per}
\end{equation}

\begin{equation}
\nu_{\phi} = \pm \frac{c^3}{2\pi GM} \frac{1}{r^{3/2} \pm a}\label{eq:kepl}
\end{equation}

\vspace{0.1in}
\hspace{-0.3in} 
where $r$ is the radius (in units of the gravitational radius $R_{\mathrm{g}}=GM/c^2$) at which the frequencies are produced,  $a$ is the dimensionless black hole spin parameter, and $M$ is the black hole mass. The signs plus and minus refer to prograde and retrograde orbits, respectively.
 
According to the RPM, type-C QPOs are associated with the Lense-Thirring precession frequency $\nu_{\mathrm{LT}}$, the lower HFQPO to the periastron precession frequency $\nu_{\mathrm{per}}$ and the upper HFQPO to the orbital frequency $\nu_{\phi}$. 
Under the hypothesis that when observed simultaneously, these frequencies are produced at the same radius $r$, the equations above - linking the black hole parameters (spin and mass) to the QPO centroid frequencies -  form a system of three equations in six variables: $(a, M, r, \nu_{\phi}, \nu_{\mathrm{per}}, \nu_{\mathrm{LT}})$.
Therefore, when a type-C QPO and two HFQPOs are detected simultaneously, one can solve the RPM system of equations exactly (\citealt{Ingram2014}), obtaining precise values for $a$, $M$ and $r$ \citep{Motta2014a,Motta2014}.

In this work we assume that the highest frequency type-C QPO detected in the HSS or ULS for each source corresponds to the Lense-Thirring precession frequency of a test particle at the ISCO. The radius of the ISCO ($r_{\rm ISCO}$), as predicted by the theory of General Relativity, is only a function of the black hole dimensionless spin parameter  (see, e.g., \citealt{Stella1998}). Therefore, substituting the expression of $r_{\rm ISCO}$ into the equations above, allows us to remove the $r$ variable from the RPM system and thus to  obtain equations that only depend on $M$ and $a$. 
This implies that, if either the mass or the spin are known, one can obtain an estimate of the other parameter detecting just one QPO and using the related RPM equation. While BH spin measurements can be obtained only indirectly and are often affected by large uncertainties, BH masses, instead, can be inferred through dynamical studies. The literature includes to date mass measurements for a few tens of compact object in binary systems, including several stellar mass BHs. We can thus make reasonable assumptions on the mass range spanned by the BHs hosted in the systems of our sample. This means that, even without a dynamically determined BH mass, we are now able to place limits on the black hole spin using the RPM.
In principle, this could be done with any QPO relevant for the RPM (type-C QPO, upper or lower HFQPO). However, type-C QPOs are much more common and easier to detect than HFQPOs and can be detected in basically all spectral/timing states \citep{Motta2016}, including the HSS and ULS.

In order to place limits on the spin in those cases where a dynamical BH mass is not available, we choose a conservative range of masses, based on the mass distribution of BHBs \citep{Casares2014}: $3-20M_{\odot}$. The smallest dynamically confirmed stellar mass black hole is GRO J1655-40, with $M = 5.4M_{\odot}$ \citep{Beer2002}.
In a similar way, the upper limit is based on the fact that the heaviest dynamically confirmed stellar mass black hole in a Galactic binary system is the one in GRS J1915+105 \citep{Reid2014}, with a black hole of $12.4M_{\odot}$. There is no evidence for heavier stellar mass black holes (even though a binary with two $30M_{\odot}$ BHs has been recently detected \citeauthor{Abbott2016} \citeyear{Abbott2016}), therefore we assumed a conservative upper limit on the mass of  $20M_{\odot}$. 

We explore the range between $3$ and $\sim150$ Hz for the LT frequency. The lower limit is chosen based on the observational evidence that all BHBs show type-C QPOs reaching at least 3 Hz \citep{Motta2015}. Even if type-C QPOs are typically observed below $\sim$30 Hz, we decided to conservatively explore a wider range of frequencies. 
Indeed, it is worth noticing that a type-C QPO produced around an almost maximally spinning black hole (e.g. $a=0.98$ obtained for GRO J1655-40 using spectroscopic methods) should have a frequency above $100$ Hz. Since in principle a type-C QPO these high could be detected, we choose a wider range of frequencies to explore this possibility and in order to guarantee an unbiased analysis.

GRS J1915+105 shows HFQPOs around $67$ Hz (\citet{Morgan1997}) while the other sources show this type of QPOs above $150$ Hz. Therefore, for periastron precession frequencies (and orbital frequency, coincident with the periastron precession frequency at the ISCO) we assume a lower limit of $50$ Hz and an upper limit of $350$ Hz since this is the highest observed HFQPO so far \citep{Strohmayer2001,Belloni2012}. 
We stress that, according to the RPM, it is possible to distinguish type-C QPOs from the HFQPO associated to either orbital or periastron motion since there is no mass-spin combination that gives the same frequency for both motions.

\begin{figure}
\includegraphics[width=0.5\textwidth]{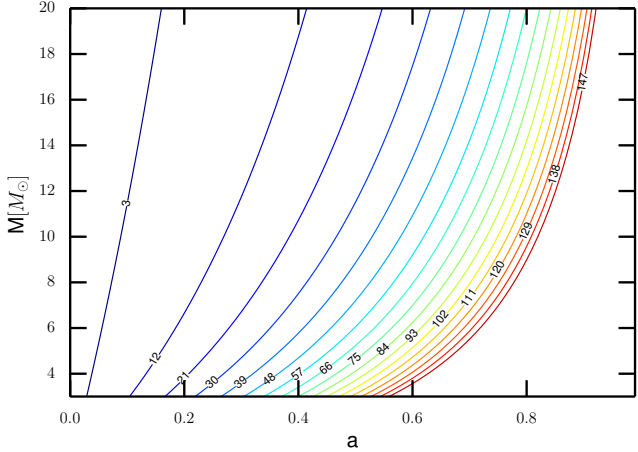}
\caption{\label{fig:nodal_spin} Couples of mass-spin values that give the LT precession frequency. The numbers are the frequencies evaluated in Hz. Each line is obtained numerically solving eq. \ref{eq:nod} assuming a frequency range from $3$ to $147$ Hz.}\label{fig:LT_plot}
\end{figure}

\begin{figure}
\includegraphics[width=0.5\textwidth]{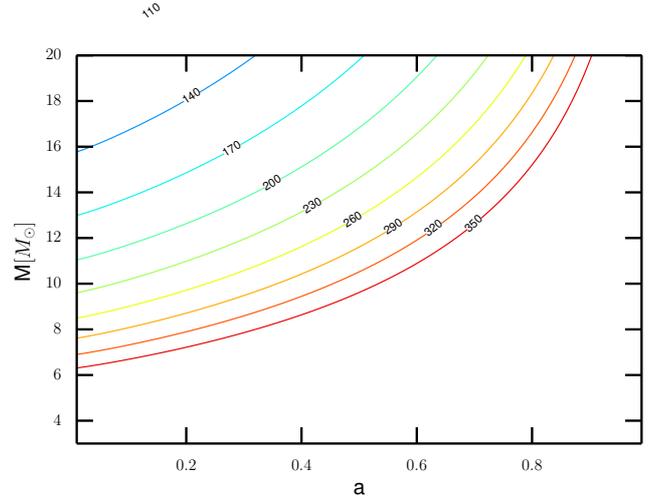}
\caption{\label{fig:per_spin} Couples of mass-spin values that give the periastron precession frequency. The exactly same plot is obtained considering the orbital frequency at the ISCO. Indeed, at the ISCO there is no radial oscillation in the frequency, thus $\nu_{\rm per} = \nu_{\phi}$. The numbers are the frequencies evaluated in Hz. Each line is obtained numerically solving eq. \ref{eq:per} assuming a frequency range from $50$ to $350$ Hz.} 
\label{fig:per_plot}
\end{figure}

In Figure \ref{fig:LT_plot} we show the Lense-Thirring precession frequency $\nu_{\mathrm{LT}}$ evaluated at $r_{\mathrm{ISCO}}$ for every mass and spin couple within the ranges $3M_{\odot} < M < 20M_{\odot}$ and $0<a<1$ (i.e. we considered only prograde orbits). Note that for clarity we only show Lense-Thirring frequencies spaced by 9 Hz. 
For comparison, we also show in Fig. \ref{fig:per_plot} the periastron precession frequencies at $r_{\mathrm{ISCO}}$ with the same spin and mass ranges. 

\bigskip

In order to place limits on the spin value, we selected the highest frequency type-C QPO for each source (considering both the QPOs we found in this work and the QPOs reported in \citealt{Motta2012,Motta2014,Motta2014a,Motta2015}) and then solved eq. (\ref{eq:nod}) for the spin $a$ after substituting the expression of $r_{\rm ISCO}$. We used either a known value for $M$ (when available in the literature), or we assumed the mass to lie in the range given above.

\section{Results}\label{sec:results}

\subsection{QPO Classification}\label{class}

Plotting the total integrated fractional rms versus the centroid frequency of a QPO is a useful method for distinguishing different types of QPOs, which in a rms vs frequency plot are known to generally form different, well defined groups (see e.g. \citealt{Casella2005}, \citealt{Kalamkar2011}, \citealt{Motta2012}). In particular, type-C QPOs are known to correlate well with the rms, forming in a frequency versus rms plane a curved track with rms decreasing for increasing frequency, breaking at $\nu_{\rm Break} \sim$10 Hz. The slope is always  steeper for the high-frequency half of the correlation track (see, e.g., \citealt{Motta2012} for the case of GRO J1655-40). HFQPOs, instead, form a different group of points at significantly higher frequencies and rms comparable to that of type-C QPOs found in the soft state (Motta et al. 2016, in prep).

\begin{figure*}
\includegraphics[width=1.0\textwidth]{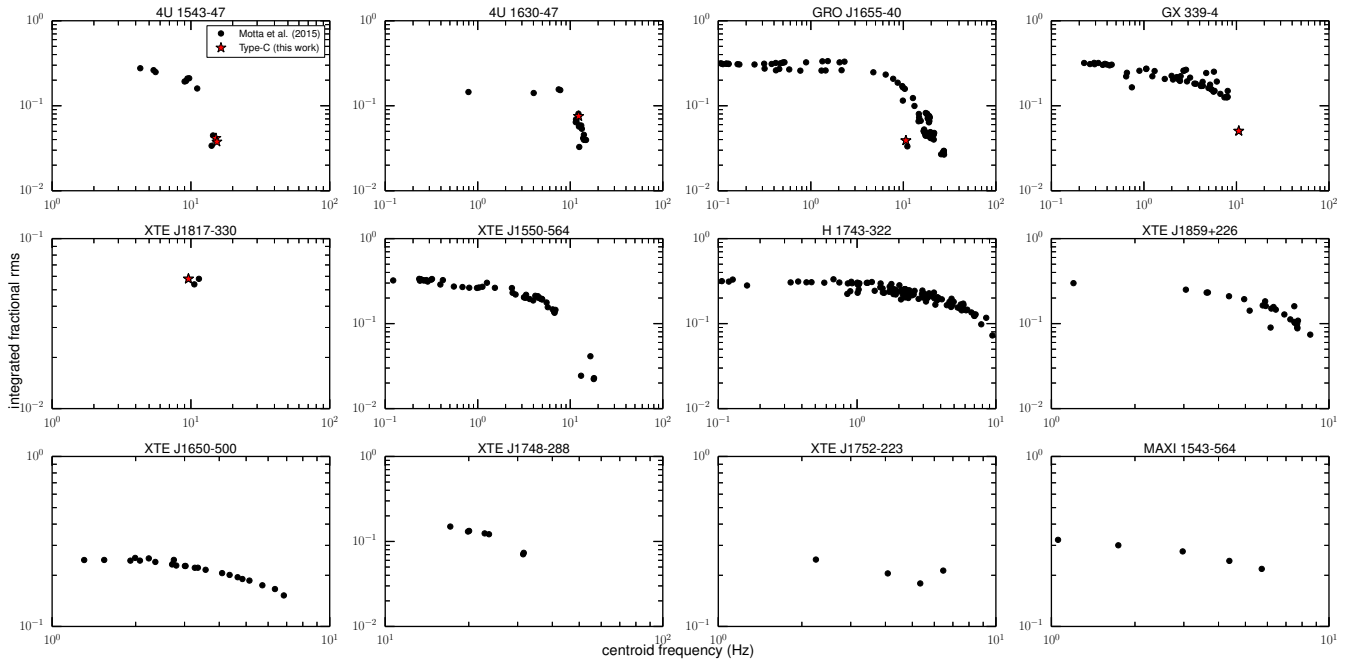} 
\caption{Integrated fractional rms versus QPO centroid frequency. Each point corresponds to a single {\it RXTE} observation. Each panel refers to a single source of our sample. }
\label{fig:rms_vs_freq}
\end{figure*}

In Fig. \ref{fig:rms_vs_freq}, we plot the integrated fractional rms versus the centroid frequency of the QPOs for each source of our sample. We see that for most of the sources the QPO centroid frequency correlates well with the rms, forming a curved track. 
In the plot we mark with black dots the QPOs reported in \citet{Motta2012,Motta2014,Motta2014a,Motta2015} and with red stars the type-C QPOs found in this work.

\subsection{Evidence of QPOs in the soft state}

It has been shown that for two sources in our sample, GRO J1655-40 and XTE J1550-564 (see \citealt{Saito2007} for GRO J1655-40 and \citealt{Kubota2004} for XTE J1550-564), the accretion disc inner radius remains constant for most of the HSS, providing a strong evidence for the accretion disc having reached the ISCO (see also \citealt{Steiner2011}, \citealt{Motta2014}, \citealt{Plant2014}). 
It is worth noticing that the QPOs in the HSS of these two sources stand out in terms of very low rms and high Q factor (see Table \ref{tab:lower_limits}).
Given the behavior of the sources of our sample (and of transient BH X-ray binaries in general) in the HIDs, it is reasonable to assume that the inner disc radius does indeed reach the ISCO any time a source is observed in the HSS, which is basically the same assumption used for the spin measurements based on spectroscopy (see e.g. \citealt{McClintock2014} and references therein). Since the type-C QPOs detected in the ULS have frequencies very close to those observed in the HSS (see e.g. \citealt{Motta2012}), we extended this assumption also to the ULS. However we decided not to place upper limits using type-C QPOs in the ULS since they might be underestimated. We describe below in Section \ref{sec:constraining} the uncertainty in the spin estimate produced by using this assumption. 

If the highest type-C QPO is detected in the HSS, we can therefore assume that it was produced at the ISCO and thus place a lower and an upper limit on the spin assuming a lower and an upper limit for the mass.
For those sources for which we did not detect a type-C QPO in the soft state, we can still use the highest frequency type-C QPO from \cite{Motta2012, Motta2014,Motta2014a,Motta2015} to place a conservative lower limit assuming a lower limit for the mass since these QPOs probably corresponded to radii larger than the ISCO. In these cases an upper limit on the spin would be meaningless (and in fact not correct) given our assumptions.

\begin{figure}
\begin{center}
\includegraphics[width= 0.32\textwidth]{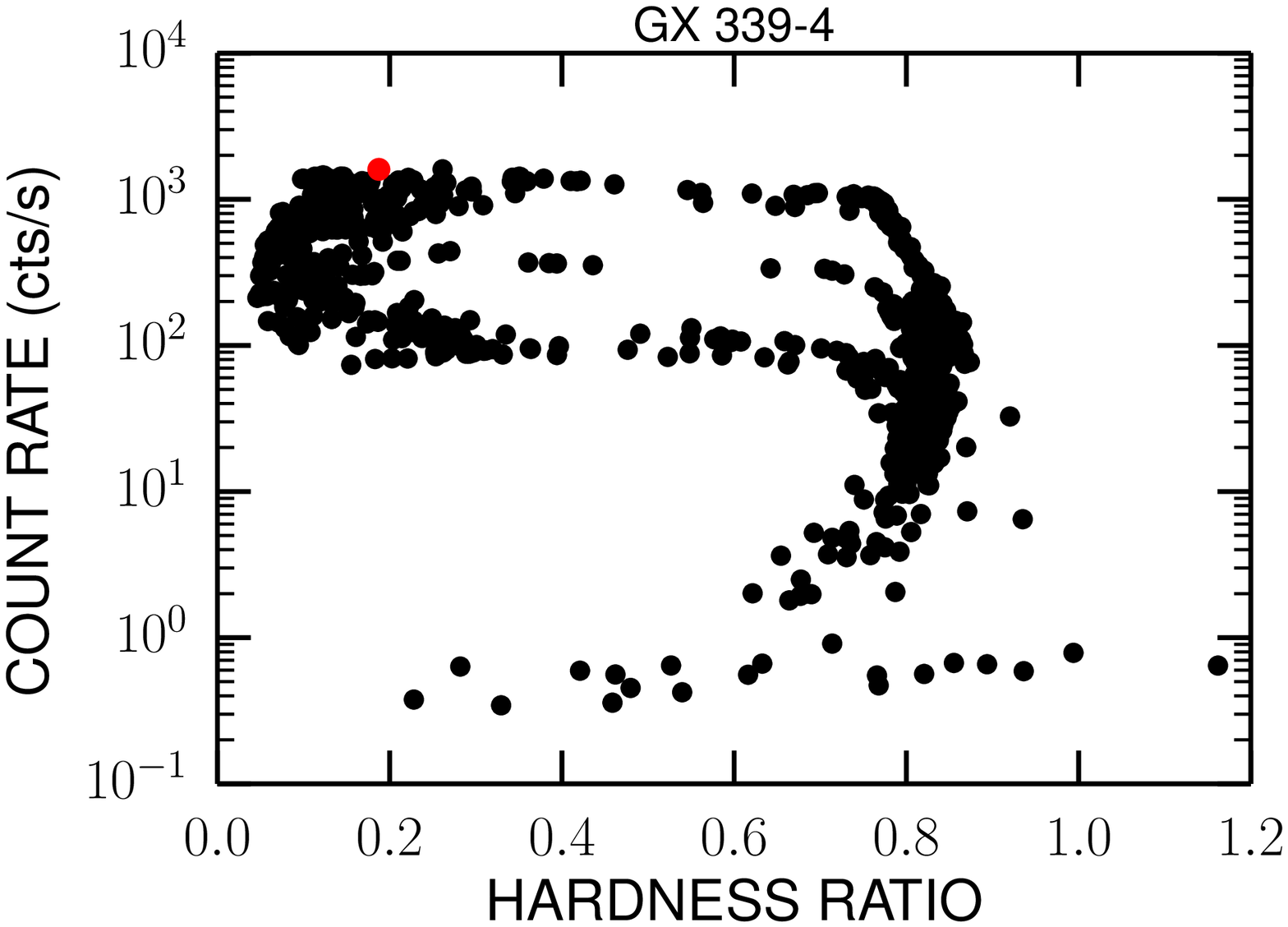}
\includegraphics[width= 0.32\textwidth]{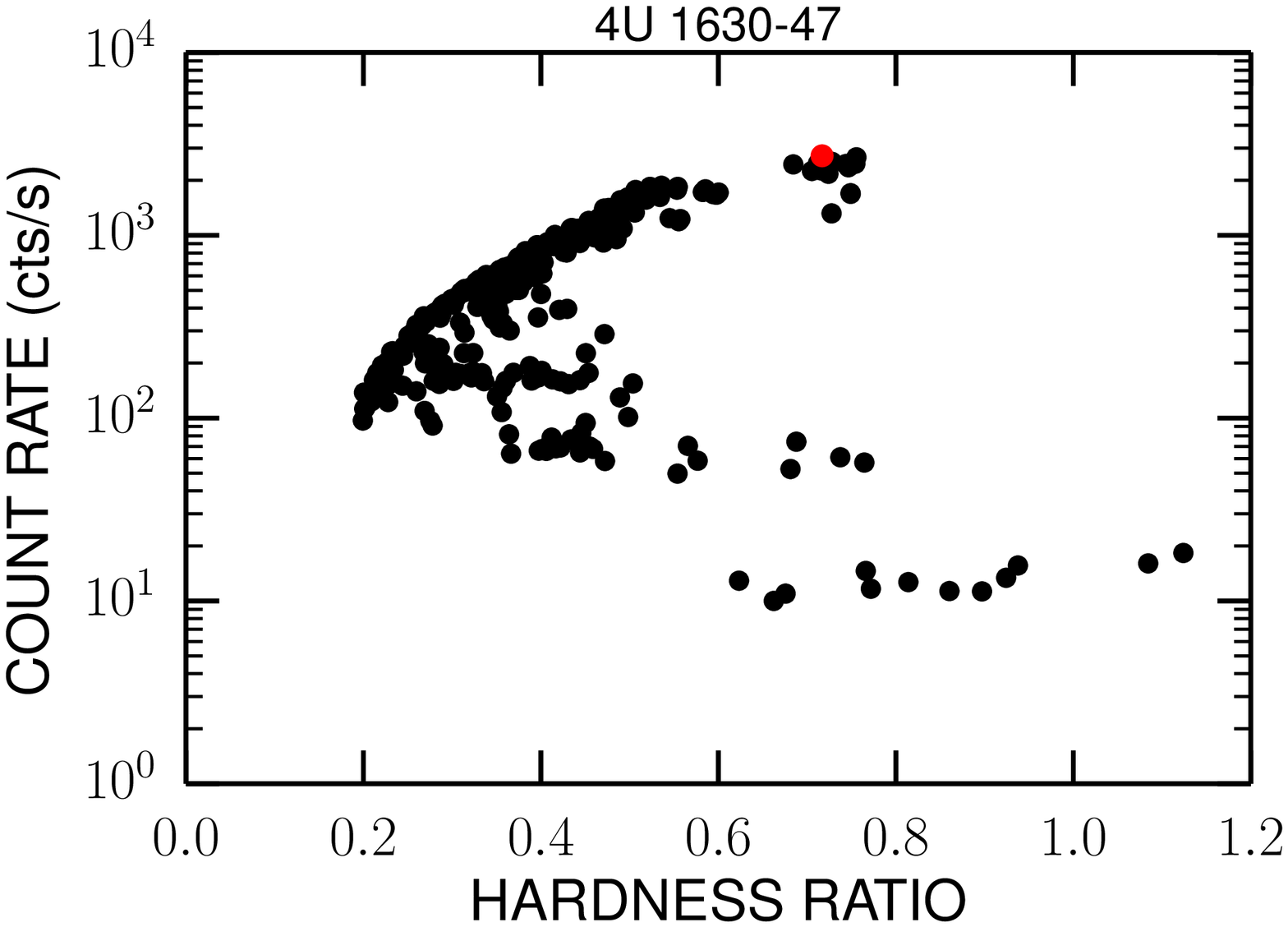}
\includegraphics[width= 0.32\textwidth]{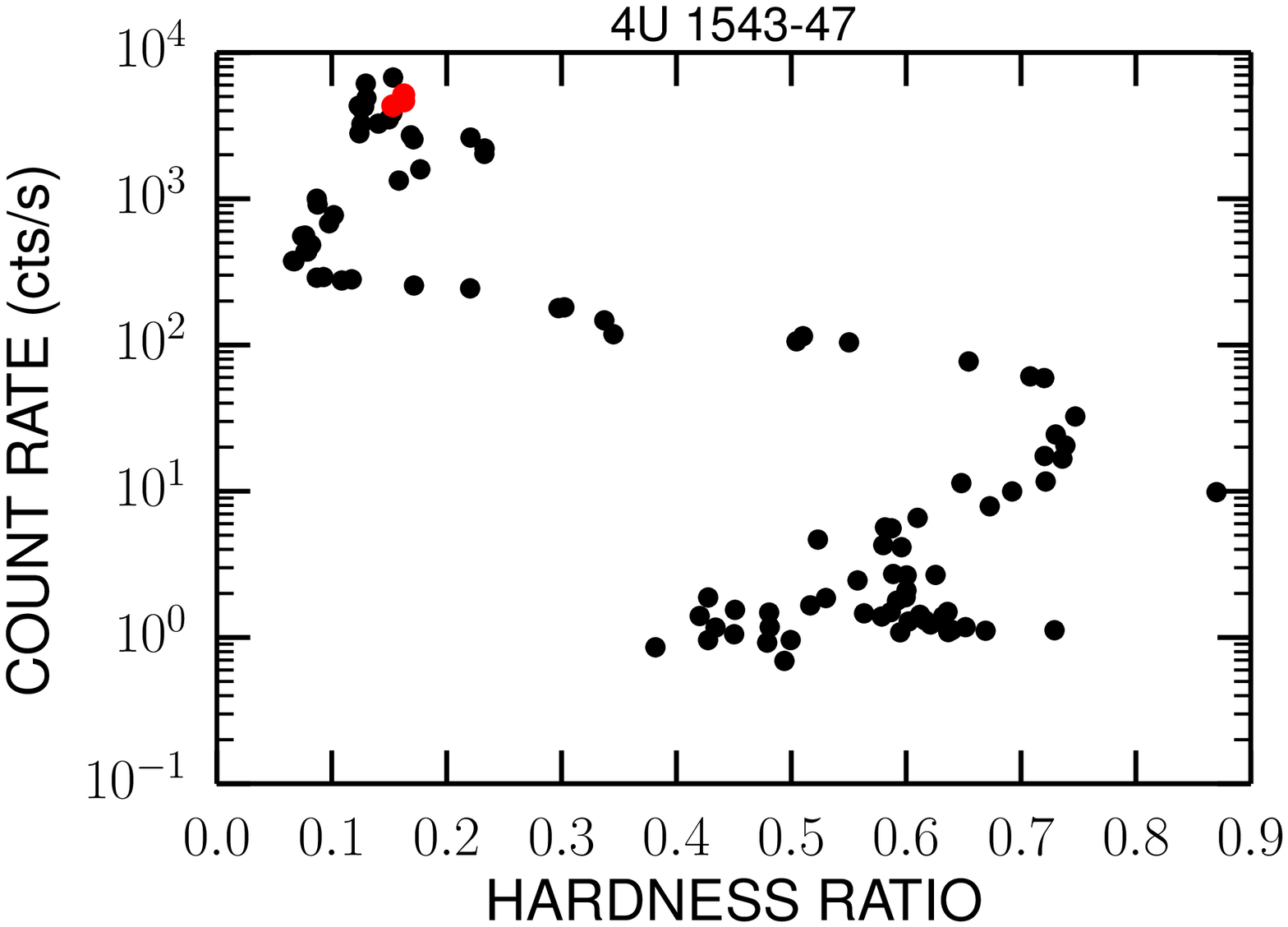}
\includegraphics[width= 0.32\textwidth]{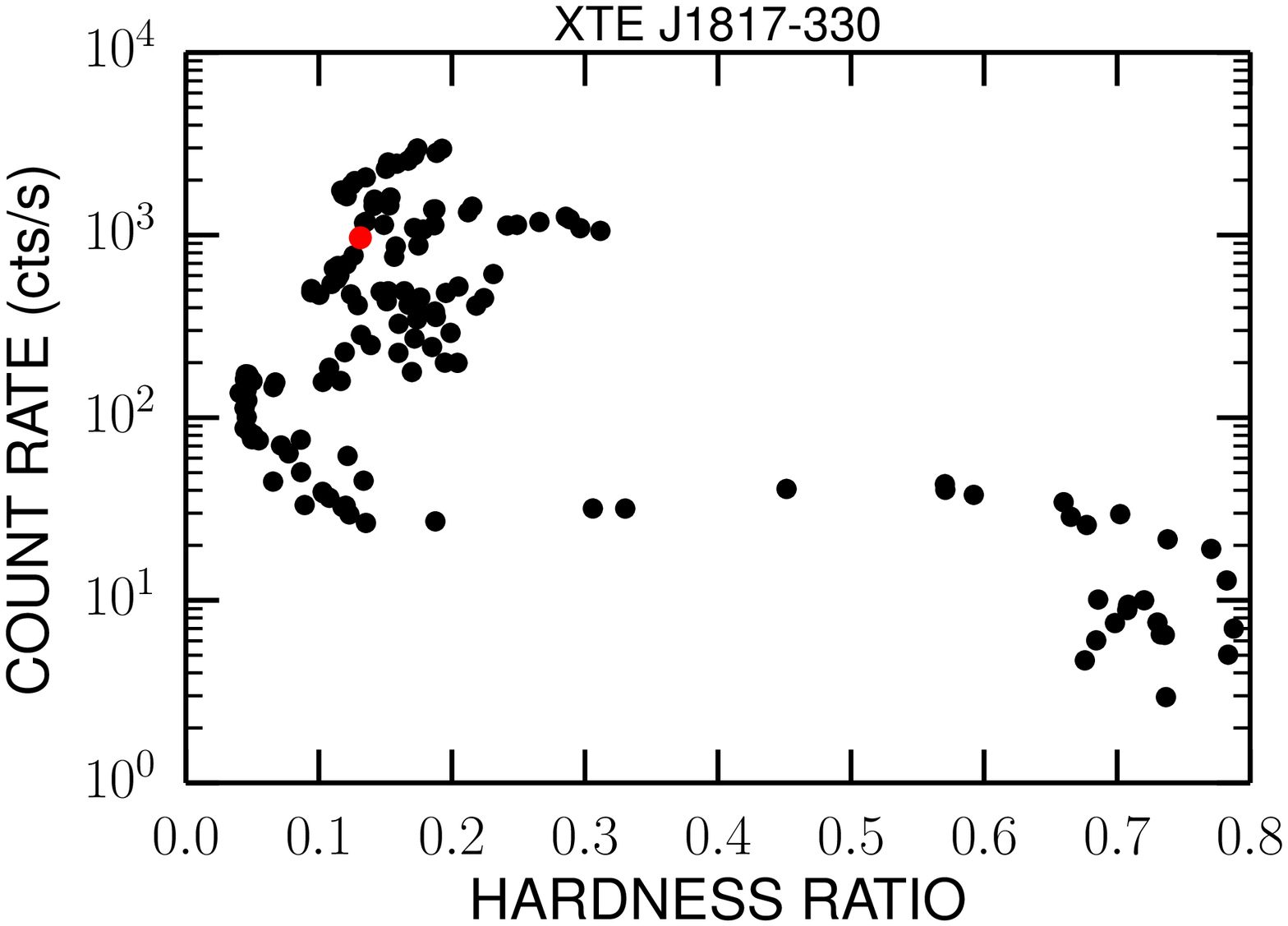}
\includegraphics[width= 0.32\textwidth]{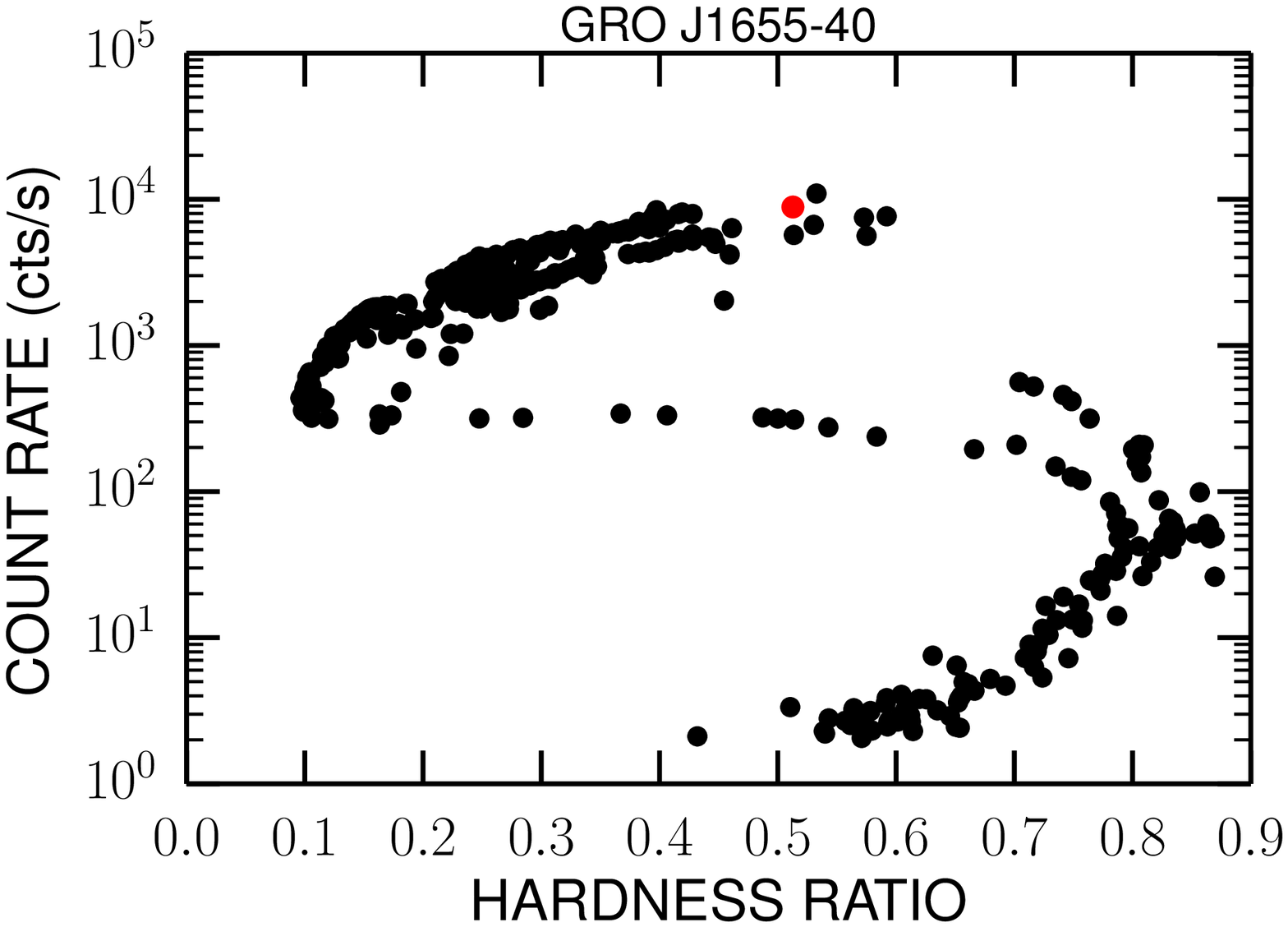}
\end{center}
\caption{HID of the $5$ sources with type-C QPOs in the HSS (GX 339-4, 4U 1630-47, 4U 1543-47, XTE J1817-330, GRO J1655-40). The red dots represent the observations that contain the type-C QPOs we found in this work. Note that the QPOs in GX 339-4, 4U 1543-47 and XTE J1817-330 were detected in soft states, while the ones in GRO J1655-40 and 4U 1630-47 were detected in the ULS. In the HID of 4U 1543-47 we marked the two different type-C QPOs we found with red dots.}
\label{fig:HID}
\end{figure}

Fig. \ref{fig:HID} contains the HIDs of the $5$ sources that showed type-C QPOs in the HSS. The red dots represent the observations of the newly detected QPOs. By looking at these diagrams and the rms values in table \ref{tab:qpos} it is clear that the type-C QPOs in GX 339-4, 4U 1543-47 and XTE J1817-330 were detected in soft states.
Instead the type-C QPOs in GRO J1655-40 and 4U 1630-47 have been detected in the ULS. We can reasonably assume that also in this state the accretion disc inner radius reaches the ISCO since the frequencies of type-C QPOs in the ULS are similar to those detected in the HSS.
In the 4U 1543-47 HID we marked with red dots the two different type-C QPOs we detected that slightly overlap with each other. Thus these HIDs contain all the $6$ new QPOs found in this work.

\subsection{Constraining the spin} 
\label{sec:constraining}

We placed lower limits on the BH spin for each source of our sample as described in Section \ref{sec:method}. For the sources with the highest type-C QPO detected in the soft state we did also place an upper limit (assuming an upper limit on the mass).
We summarize the results obtained applying the RPM to our sample in Table \ref{tab:lower_limits}. In the first column are listed the source names and the second contains the Obs-ID of the highest detected type-C QPO. The third, fourth and fifth columns contain the highest type-C QPO frequency and quality factor and the total fractional rms respectively. The sixth column indicates whether the highest type-C QPO was detected in this or in previous works \citep{Motta2012,Motta2014,Motta2014a,Motta2015,Homan2005b}. The last two columns contain the state in which the highest type-C QPO was detected and the spin limits obtained respectively.

We would like to stress that we performed a systematic search for type-C QPOs in the HSS and ULS. For those sources for which we did not detect any type-C QPO in these states we used the type-C QPOs reported in \cite{Motta2015}.
For completeness we were made aware of the fact that there are two QPOs at even higher frequencies missing in \cite{Motta2015} for XTE J1859+226 and XTE J1650-500. Thus our lower limits for these two sources are slightly underestimated. We decided not to add these QPOs in our work since we do not want to bias the analysis performed. 

For most of the sources of our sample there is no independent mass measurement, thus we assumed the mass to lie in the range $3-20M_{\odot}$. 
In a few cases, however, we have better constraints on the black hole mass and thus on the spin value.
The lower limit for the black hole mass in GX 339-4 is $6M_{\odot}$ \citep{Hynes2003,Munoz-Darias2008}. Thus we can place a stricter lower limit on the spin value: $a_{\rm lower} = 0.16$. Since we are assuming BH masses up to $20M_{\odot}$, the upper limit for the spin of GX 339-4 is $a_{\rm upper} = 0.38$.

For both GRO J1655-40 ($5.31\pm0.07\,M_{\odot}$ \citealt{Motta2014}) and XTE J1550-564 ($9.1\pm 0.6M_{\odot}$ \citealt{Orosz2011}) we have a good independent mass measurement. Using these dynamically estimated mass values, we can narrow significantly the spin range of values for the two sources (see Table \ref{tab:lower_limits}). The spin values found by \cite{Motta2014,Motta2014a} ($a = 0.290\pm0.003$ for GRO J1655-40 and $a = 0.34\pm0.01$ for XTE J1550-564) are, as expected,  in good agreement with the range we inferred using the RPM with the highest type-C QPO at the ISCO.

\begin{table*}
\centering
\caption{Spin limits inferred from the highest frequency type-C QPO at ISCO according to the RPM. Both limits are evaluated assuming a black hole mass in the range $3-20M_{\odot}$. In the case of GX339-4 the lower limit corresponds to $6M_{\odot}$. For XTE J1550-564 and GRO J1655-40 the limits correspond to the measured masses $9.1\pm 0.6M_{\odot}$ and $5.31\pm 0.07M_{\odot}$ respectively \citep{tmd2010b,Motta2014a,Motta2014}. The first column contains the name of the source, the second is the Obs-ID of the highest type-C QPO detected. The third, fourth and fifth columns contain the properties of the QPO, i.e. frequency and the Q factor, and the total fractional rms. The sixth column indicates whether the highest type-C QPO has been detected in this work or the previous ones. The seventh column indicates the state in which the QPO was detected and the last column contains the spin limits.}
 \label{tab:lower_limits}
\begin{tabular}{lcccccccc}
  \hline
  Target & Obs-ID & $\nu_{\mathrm{max}}$ (Hz) & Q & rms & New ? & state & spin limits\\
  \hline
  GX 339-4 & 92085-01-02-03 & $10.59 \pm 0.18$ & $3.46\pm0.50$ & $5.05\pm0.03$ & yes & HSS &0.16 - 0.38 \\
  4U 1630-47 & 80117-01-07-01 & $14.80\pm0.28$ & $2.17\pm0.27$ & $9.5\pm0.4$ & no & ULS & >0.12   \\
  4U 1543-47 & 70133-01-01-00 & $15.37 \pm 0.18$ & $2.57 \pm 0.27$ & $4.2\pm0.03$ & yes & HSS & 0.13 - 0.47 \\
  XTE J1859+226 & 40124-01-14-00 & $8.56 \pm 0.06$ & $2.76\pm0.18$ & $12.4 \pm 0.4$ & no & HIMS & >0.07   \\
  XTE J1650-500 & 60113-01-13-02 & $6.84\pm0.05$ & $8.05\pm1.23$ & $20.8\pm0.3$ & no & HIMS & >0.06   \\
  XTE J1817-330 & 91110-02-32-00 & $9.6 \pm 0.5$ & $2.93 \pm 0.82$ & $5.7\pm0.1$ & yes & HSS & 0.08 - 0.36  \\
  XTE J1748-288 & 30171-02-01-00 & $31.55 \pm 0.13$ & $6.00\pm0.42$ & $10.2 \pm 0.5$ & no & HIMS & >0.23  \\
  XTE J1752-223 & 95360-01-11-00 & $6.46 \pm 0.13$ & $3.6\pm1.1$ & $23.4 \pm 1.2$ & no & HIMS &  >0.06 \\
  XTE J1550-564 & 40401-01-48-00 & $18.10 \pm 0.06$ & $19\pm4$ & $6.2 \pm 0.1$ & no & HSS & 0.31 - 0.34 \\
  MAXI J1543-564 & 96371-02-02-01 & $5.72 \pm 0.04$ & $10.2\pm1.4$ & $27.3 \pm 1.2$ & no & HIMS & >0.05 \\
  H1743-322 & 80135-02-03-00 & $14.6 \pm 0.2$ & $12.6\pm6.3$ & $4.4 \pm 0.2$ & no & ULS & >0.12   \\
  GRO J1655-40 & 91702-01-17-01 & $27.51 \pm 0.13$ & $44\pm21$ & $2.7 \pm 0.1$ & no & HSS & 0.29 - 0.31 \\
 \hline
 \end{tabular}
\end{table*}

\begin{figure}
\centering
\includegraphics[width=0.4\textwidth] {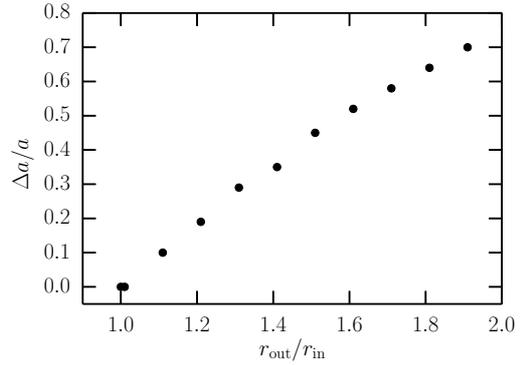}
\caption{Relative change in the spin estimate assuming that the QPO is produced by an extended rigidly precessing structure \citep{Ingram2009} rather than a test particle at ISCO, as a function of the radial extent of the precessing structure $r_{\rm out}/r_{\rm in}$, where $r_{\rm out}$ is the outer radius and $r_{\rm in}=r_{\rm ISCO}$ is the inner radius of the structure.}
\label{fig:spinerror}
\end{figure}

As mentioned above, there might still be the possibility that, even in the soft state, the disc does not extend down to the ISCO, but somewhat close to it. In this case, one can still relate the QPO frequency to the spin of the black hole, because we expect that the hot inner flow would have a finite and small radial extent so that it would precess as a rigid body \citep{Ingram2011,Franchini2016}. Thus, we can assume that the hot inner flow extends from the ISCO to an outer radius $R_{\rm out}$, and then compute the spin value that would reproduce a given QPO frequency using, for example eq. (1) in \citet{Ingram2011}. In this way, we can thus estimate the associated change to the spin value obtained from the RPM model, as a function of $R_{\rm out}$. For instance, using the highest type-C QPO detected in GRO J1655-40 (i.e. $27.51$ Hz) together with the mass estimate for this source, we can evaluate the uncertainty with respect to the value obtained using the RPM. The results in this case are shown in Figure \ref{fig:spinerror}. The $x$-axis is the ratio between the outer radius of the inner accretion flow and the ISCO and the $y$-axis is the relative change in the spin value.
A similar result holds also for different QPO frequencies.

\section{Discussion}\label{sec:discussion}

We analyzed {\it RXTE} observations of $12$ black hole transients that showed a transition to the soft state. We considered observations in the HSS and ULS of these sources.
We then assumed that the highest frequency type-C QPO has been produced at the ISCO and we applied the RPM in order to place limits on the spin value. This is done either using a known value of the black hole mass or assuming the mass in a reasonable range of values, decided \textit{a priori} ($3-20M_{\odot}$ in our case).

This method is based on the assumption that the inner disc truncation radius is constant and coincident with ISCO when a BHB is observed in the HSS or ULS.
Demonstrating the coincidence of the inner disc truncation radius with ISCO is problematic and involves spectroscopic measurements that are generally affected by large uncertainties. Therefore, our assumption constitutes a relevant caveat. However, both observations and simulations provide strong evidence about the stability of the inner disc radius over time in disc-dominated states of BHBs (e.g. \citealt{McClintock2014}). \cite{Tanaka1995} showed that the inner radius is remarkably stable also if the thermal flux of
the source steadily decreases on timescales of months by one or two orders of magnitude. Similar evidence for a constant inner disc radius in the soft state has been demonstrated for many sources (\citealt{Done2007} and references therein, \citealt{Plant2014}, \citealt{Plant2015}).
According to \cite{Tanaka1995}, the constancy of the inner radius suggests that it is related to the ISCO. More recently \cite{Steiner2010} found another evidence of a non varying disc inner radius over time for the source LMC X-3, unaffected by the high variability of this source.
Nevertheless, since we cannot be certain of the coincidence of the inner disc radius with ISCO for all the sources of our sample, excluded GRO J1655-40 and XTE J1550-564 (for which the coincidence has been demonstrated explicitly), we advice the reader to consider the inferred limits with caution. 

The rigid precession of the inner accretion flow as an explanation of type-C QPOs has been supported by several studies \citep{Ingram2009,Ingram2014,Ingram2016}. The rigid precession model tends asymptotically to the RPM for radii approaching the ISCO, therefore, given the stability of the disc inner radius in the soft state, the RPM represents a good approximation of the rigid precession for truncation radii (hence hot flow \textit{outer} radii) very close to the ISCO \citep{Ingram2009}.

The approach we used in this work differs from that of \cite{Ingram2014}, which always requires that at least two QPOs are detected simultaneously (the type-C QPO and one HFQPO) in order to place an upper or lower limit on spin and mass, depending on the case considered. 
Here we assumed that the highest frequency type-C QPO is produced at the ISCO.  Thus the RPM system of equations is reduced to just one equation (i.e. eq. (\ref{eq:nod})) that gives either a lower limit (if the highest type-C QPO was not detected in the HSS) or a range of spin values. The ranges are broad if there is no independent (e.g. dynamical) mass measurement available, or significantly narrower if the mass is known. The reason is that the uncertainty on the spin is completely dominated by the error on the mass (much larger than the error on the QPO centroid frequency).

In principle we can never be completely sure that the type-C QPO we assumed to be the highest is actually the highest produced by the source in the soft state. There is always the possibility that an even higher frequency type-C QPO has been produced and not detected. Using a type-C QPO that is not at the highest possible frequency for a given source results in an underestimation of both upper and lower limit of the spin. In other words, when estimating the spin range from a type-C QPO that is not the highest in frequency, we would be following the wrong track in Fig. \ref{fig:nodal_spin}, that would be found to the left of the track corresponding to the actual highest frequency type-C QPO. This effect is larger for QPOs that are not detected in the HSS, therefore produced at a radius that is almost certainly larger than ISCO.  

We note that the different tracks in Fig. \ref{fig:nodal_spin} tend to lay closer to each other for high frequencies (i.e. above $80$ Hz). Thus, the error on the spin estimate might be dominated by that on the mass if the QPO frequency is very high. In general, the two uncertainty that contribute to the error on the spin value are that on the black hole mass and the identification of the highest type-C QPO. For instance, if the actual highest type-C QPO was produced at, say, $30$ Hz instead of being detected at $21$ Hz, the error that we commit on the spin estimate is of the order of a few percent. 

\smallskip

\subsection{Comparison with other methods}

\begin{figure}
\includegraphics[width=0.5\textwidth] {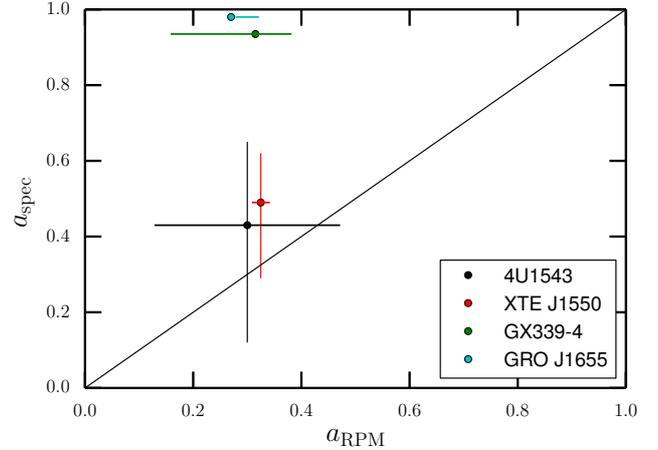}
\caption{The horizontal lines represent the different ranges of spin values obtained in this work for each source. The black, red, green and cyan lines refer to 4U1543, XTE J1550, GX 339-4 and GRO J1655 respectively. On the $y$-axis the dots represent the values obtained using spectroscopic methods with the correspondent error bars. The black line represents the case $a_{\rm spec} = a_{\rm RPM}$.}
\label{fig:spin_compar}
\end{figure}

For some sources of our sample the spin has been also measured through X-ray spectroscopy by fitting the thermal disc emission or by modelling the disc reflection spectrum. Fig. 
\ref{fig:spin_compar} shows the spin measurements found in the literature obtained from X-ray spectroscopy versus the spin values we inferred from timing analysis in this work. The black line marks the position where the measurements should fall if there was perfect agreement between spectroscopy based measurements and timing based measurements.  In two out of five sources (XTE J1550-564 and 4U 1543-47) the spin measurements obtained with spectroscopy and timing do agree. 
In particular, \cite{Morningstar2014} obtained a spin value $a = 0.43^{+0.22}_{-0.31}$, in agreement with our values, for 4U 1543-47 using both the spectral continuum fitting method and the iron line fitting method simultaneously. \cite{Steiner2011} used data from the BHB XTE J1550-564 to model the continuum and to fit the iron line obtaining two different spin measurements, that combined resulted in a spin of $a = 0.49^{+0.13}_{-0.20}$ which is marginally consistent with the spin range that we report (see Table \ref{tab:lower_limits}). 

For 4U 1630-47 \cite{KingWaltonMiller2014} measured the spin by fitting the continuum spectrum of the source, obtaining $a = 0.985^{+0.005}_{-0.014}$, which is in principle compatible with our lower limit (i.e. $0.12$). 
For XTE J1752-223 and GX 339-4, spectroscopy returned  high spin values using both continuum fitting and the reflection model \citep{Reis2011,Miller2002,Reis2008}. \cite{Reis2011} found an intermediate black hole spin $a = 0.52 \pm 0.11$ for XTE J1752-223 while for GX 339-4 an almost extreme spin value $a = 0.935\pm 0.010$ has been inferred.  
While for XTE J1752-223 we were able to place only a lower limit on the spin thus our findings are still consistent with the results obtained using spectroscopy, the spectroscopic estimate of the spin in GX 339-4 is not consistent with the range obtained using the RPM.

\cite{Motta2014} measured with unprecedented precision the mass and spin of the BH in GRO J1655-40 applying the RPM using the three QPOs that have been simultaneously observed \citep{Motta2012,Strohmayer2001}. They inferred a mass $M = 5.31 \pm 0.07M_{\odot}$ and a spin value $a = 0.290 \pm 0.003$ that is, as expected, in good agreement with the range obtained in this paper.  Through spectroscopic studies, different, much higher, values of the spin have been inferred. 
By modelling the thermal spectral continuum \cite{Shafee2006} found a spin range $a = 0.65-0.75$. \cite{Reis2009} placed a lower limit $a=0.9$ by fitting the strong reflection features in the spectrum and \cite{Miller2009} obtained a highly spinning black hole with $a = 0.94-0.98$ using the same method. Both the spectroscopic methods rely on the assumption that the inner edge of the disc corresponds to the ISCO, which might not be a correct assumption in all spectral states of active black holes (see, e.g. \citet{Done2007}). Furthermore, in order to use the spectral continuum fitting method, one must also know the black hole mass, the inclination of the (inner) accretion disc with respect to the line of sight and the distance to the source. All these factors introduce large uncertainties and possibly significant biases in the spin measurement. For instance, as already noted by \cite{Motta2014a} in the case of GRO J1655-40, the disagreement between spin measurements inferred through different methods might be due to the high inclination of the inner accretion disc with respect to the outer disc (i.e. the orbital inclination) and to the line of sight.  
When such misalignment is not taken into account, the final spin measurement can be largely biased. 
The RPM method does not depend on any of these variables, and in the case where 3 QPOs are simultaneously detected, it allows to obtain self consistently mass and spin of the BH. The precision on the inferred parameters is only driven by the precision in the QPO frequencies measurement. 

More spectroscopy-based and timing-based measurements of the spin are necessary to confirm the validity of the RPM method and to test its consistency with other methods. New generation satellites (such as eXTP \citet{Zhang2016}) will provide more and better data to test this and other methods for spin measurements.

Finally, as already noted by \cite{Motta2014}, we would like to stress that the RPM
is a simplified model aimed at describing rather complex physical conditions, such as those encountered in the innermost disc regions around an accreting compact object. The RPM does not include yet a production mechanism for the QPO signals (but see \citealt{Psaltis2000} or \citealt{Schnittman2006} for possible mechanisms), which remains an open question to be investigated in the future.

\section{Summary and conclusions}\label{sec:concl}

In this work we placed limits on the value of the black hole spin for several binaries through the sole use of X-ray timing. In particular we looked for type-C QPOs in the soft state for a number of sources selected based on the presence of type-C QPOs in their PDS, and then applied the RPM. This choice is justified by the fact that in the HSS the disc is expected to extend down or very close to the ISCO. As a consequence, the inner accretion flow is extremely narrow in this state and we assumed that its Lense-Thirring frequency can be well described by that of a test particle at the ISCO. 
According to the RPM, we associated the highest frequency type-C QPO with the Lense-Thirring precession frequency of a test particle at the ISCO. We then inferred lower spin limits for all the sources of our sample and upper limits for the ones that showed type-C QPOs in the HSS or ULS, assuming a reasonable range of black hole masses or using the actual mass value for those source where a dynamical mass measurement is available.
We successfully applied the RPM to all the sources in our sample that showed type-C QPOs in the HSS or ULS.

Our results are in good agreement with those inferred using other spectroscopy-based methods (continuum fitting method and the K-$\alpha$ line modelling ) only in two cases. 
Next generation satellites, characterized by a higher signal-to-noise ratio might help in solving the issue of spin measurements.

\section{Acknowledgments} 

We are grateful to the referee, Dr. Jeroen Homan, that provided useful comments and suggestions that helped in significantly improve our work.
AF acknowledges the transition grant of the Universit\`a degli Studi di Milano and the University of Oxford for hospitality.
SEM acknowledges the University of Oxford and the Violette and Samuel Glasstone Research Fellowship program.

\appendix
\section{Power density spectra and properties of QPOs} 

\begin{figure}
\centering
\includegraphics[width=0.5\textwidth]{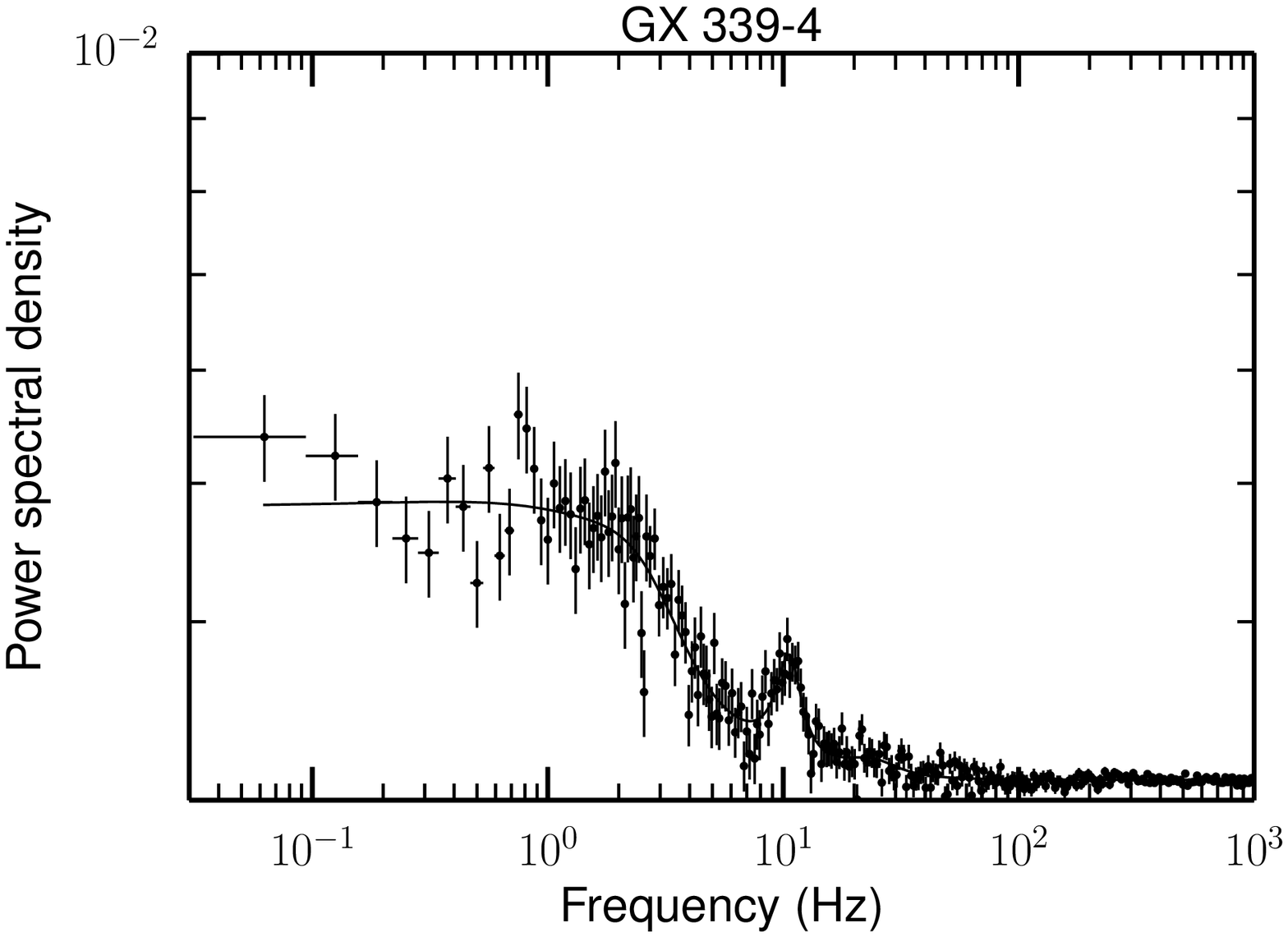}
\includegraphics[width=0.5\textwidth]{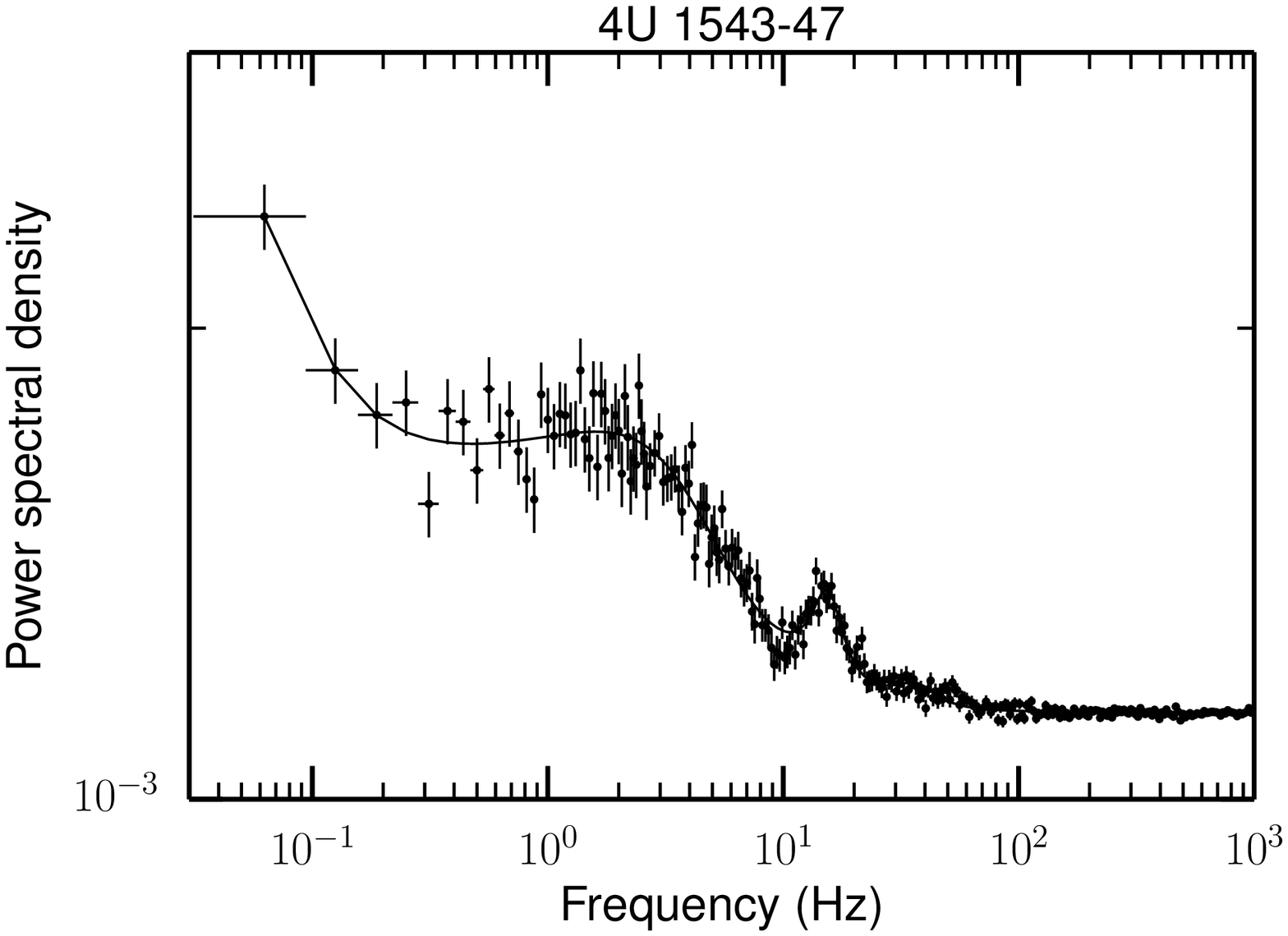}
\includegraphics[width=0.5\textwidth]{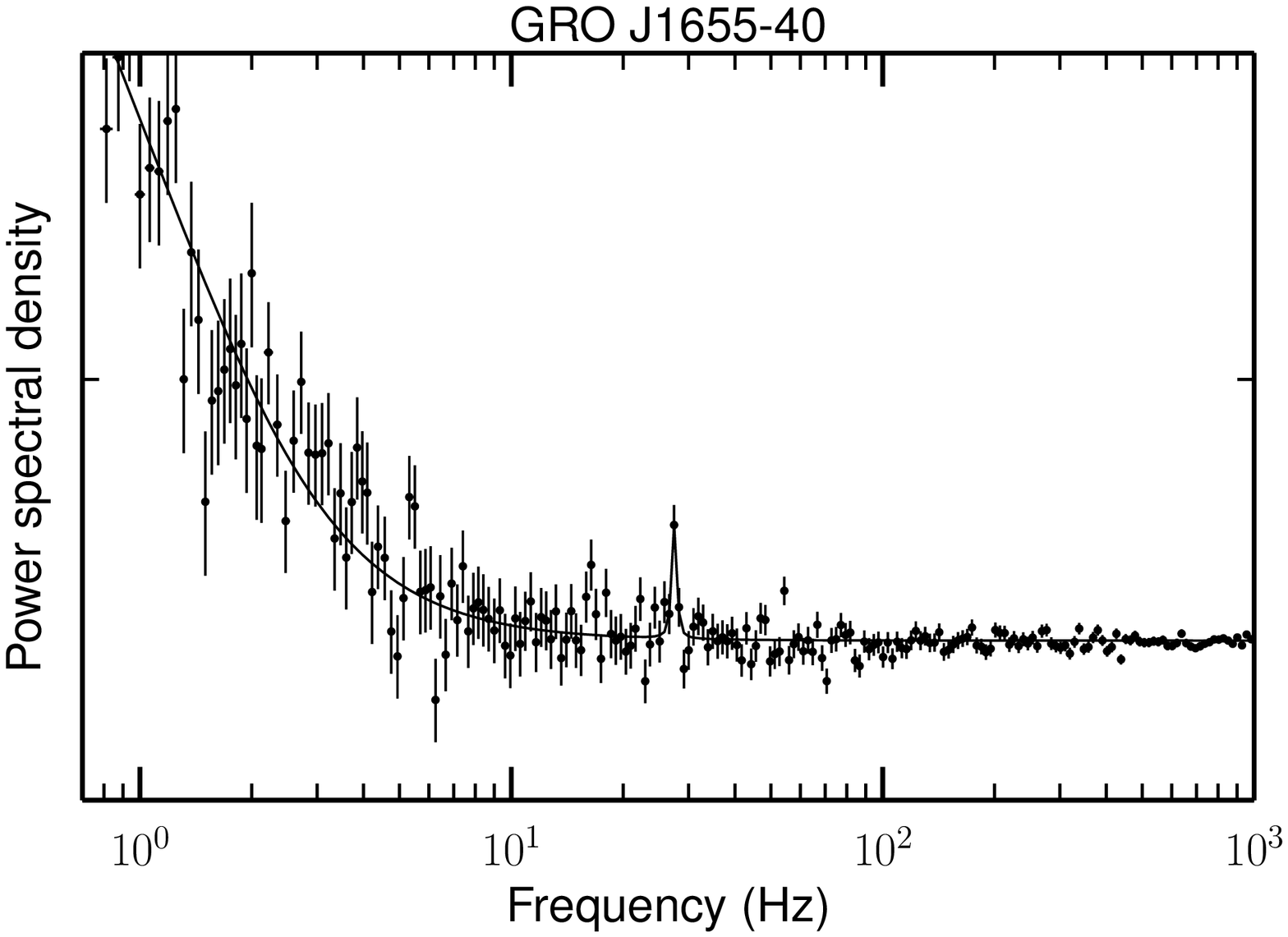}
\caption{Upper panel: PDS obtained from observation 92085-01-02-03 of the source GX 339-4 during its outburst in 2007. The type-C QPO found at 10.6 Hz is clearly visible. Middle panel: PDS obtained from observation 70133-01-01-00 of the source 4U 1543-47. The type-C QPO found at 15.4 Hz is clearly visible. Lower panel: PDS obtained from observation 91702-01-17-10 of the source GRO J1655-40. There is a small but significant peak at roughly 30 Hz. This is the highest type-C QPO detected for this source.}
\label{fig:pds}
\end{figure}

Figure \ref{fig:pds} shows a selection of PDS containing a type-C QPO detected in the HSS and in the ULS. From top to bottom, the PDS come from GX 339-4, 4U 1543-47 and GRO J1655-40 respectively.
The first two QPOs are characterized by a strong and narrow peak at frequencies of the order of $10$ Hz. The third is a small though significant (3.96$\sigma$, single trial) peak at $27.5$ Hz.

We report in Table \ref{tab:qpos} the properties of the new QPOs detected in this work: the ObsID, the centroid frequency, the quality factor and the integrated fractional rms.
Note that the total fractional rms measured for the observation where a given QPO is detected is always $\leq 5\%$, which shows that all these QPOs were detected in soft states. As noted elsewhere, this implies that the inner disc radius is close or coincident to the ISCO and the RPM can be applied to obtain a constrain on the spin of the central black hole.

\begin{table}
\centering
\begin{tabular}{llllr}
\hline
ObsID    & & $\nu$ (Hz) & Q & rms \\
\hline
\multicolumn{5}{c}{GX 339-4} \\
\hline
92085-01-02-03 & & 10.59 & 3.46 & 0.05  \\
\hline
\multicolumn{5}{c}{4U 1630-47} \\
\hline
70417-01-09-00 & & 12.3 & 2.15 & 0.033  \\
\hline
\multicolumn{5}{c}{4U 1543-47} \\
\hline
70133-01-01-00 & & 15.37 & 2.57 & 0.042  \\
70133-01-05-00 & & 15.0 & 3.75  & 0.046   \\
\hline
\multicolumn{5}{c}{XTE J1817-330} \\
\hline
91110-02-32-00 & & 9.6 & 2.93 & 0.057  \\
\hline
\multicolumn{5}{c}{GRO J1655-40} \\
\hline
10255-01-05-00 & & 10.7 & 2.7 & 0.011  \\
\hline
\end{tabular}
\caption{Properties of the QPOs found in this work divided by source. For each QPO the first column contains the ObsID, the second is the centroid frequency, the third is the quality factor Q and the last one is the integrated fractional rms.}
\label{tab:qpos}
\end{table}




\bibliographystyle{mnras}
\bibliography{biblio} 


\bsp	
\label{lastpage}
\end{document}